\documentclass{aa}  

\usepackage{xcolor}
\usepackage{graphicx}
\usepackage{amssymb}
\usepackage{amsmath} 
\usepackage{mathrsfs} 
\usepackage{stmaryrd}
\usepackage{dsfont}
\usepackage{natbib}
\usepackage{graphicx}
\usepackage[T1]{fontenc}
\usepackage{txfonts}
\usepackage{bm}
\usepackage{mathtools}
\usepackage{empheq}
\usepackage{natbib,twoopt}
\bibpunct{(}{)}{;}{a}{}{,}             %% natbib format for A&A and ApJ
\makeatletter
  \newcommandtwoopt{\citeads}[3][][]{\href{http://adsabs.harvard.edu/abs/#3}%
    {\def\hyper@linkstart##1##2{}%
     \let\hyper@linkend\@empty\citealp[#1][#2]{#3}}}
  \newcommandtwoopt{\citepads}[3][][]{\href{http://adsabs.harvard.edu/abs/#3}%
    {\def\hyper@linkstart##1##2{}%
     \let\hyper@linkend\@empty\citep[#1][#2]{#3}}}
  \newcommandtwoopt{\citetads}[3][][]{\href{http://adsabs.harvard.edu/abs/#3}%
    {\def\hyper@linkstart##1##2{}%
     \let\hyper@linkend\@empty\citet[#1][#2]{#3}}}
  \newcommandtwoopt{\citeyearads}[3][][]%
    {\href{http://adsabs.harvard.edu/abs/#3}
    {\def\hyper@linkstart##1##2{}%
     \let\hyper@linkend\@empty\citeyear[#1][#2]{#3}}}
    \renewcommand*\aa@pageof{, page \thepage{} of \pageref*{LastPage}}
\makeatother

\defcitealias{Provost}{P81}
\usepackage[breaklinks=true]{hyperref} %% to avoid \citeads line fills
\renewcommand{\vec}[1]{\if#1\relax\bm{#1}\else\mathbf{#1}\fi}

\newcommand{\difp}[2]{\frac{\partial #1}{\partial#2}}
\newcommand*\dif{\mathop{}\!\mathrm{d}}
\newcommand{\difd}[2]{\frac{\dif{#1}}{\dif{#2}}}

\begin{document} 

   \title{Rossby modes in slowly rotating stars:\\ depth dependence in distorted polytropes with uniform rotation}
   \author{C. Damiani \inst{1}
          \and
          R. H. Cameron \inst{1}
          \and
          A. C. Birch \inst{1}
          \and
          L. Gizon \inst{1, }\inst{2,}\inst{3}
          } 

   \institute{Max-Planck-Institut f{\"u}r Sonnensystemforschung, Justus-von-Liebig-Weg 3, 37077 G{\"o}ttingen, Germany\\
              \email{damiani@mps.mpg.de}
    \and
    Institut f{\"u}r Astrophysik, Georg-August-Universit{\"a}t G{\"o}ttingen, Friedrich-Hund-Platz 1, 37077 G{\"o}ttingen, Germany
   \and
   Center for Space Science, NYUAD Institute, New York University Abu Dhabi, PO Box 129188, Abu Dhabi, UAE}
              
   \date{Received  xxx ; accepted xxx}
\authorrunning{Damiani et al.}
\titlerunning{Rossby modes in slowly rotating stars}
% \abstract{}{}{}{}{} 
% 5 {} token are mandatory
 
  \abstract
   {Large-scale Rossby waves have recently been discovered from measurements of horizontal surface and near-surface solar flows \citepads{2018NatAs...2..568L}. }
   {We are interested in understanding why only the sectoral modes are seen in the observations and also in modelling the radial structure of the observed modes. To do so, we characterise here the radial eigenfunctions of r modes for slowly-rotating polytropes in uniform rotation.}
   {We follow \citet{Provost} and consider a linear perturbation theory to describe nearly-toroidal stellar adiabatic oscillations in the inviscid case. We use perturbation theory to approximate the solutions to the fourth order in the rotational frequency of the star. We numerically solve the eigenvalue problem, concentrating on the behaviour where the stratification is nearly adiabatic.}
   {We find that for free-surface boundary conditions on a spheroid of non-vanishing surface density, r modes can only exist for $\ell=m$ spherical harmonics in the inviscid case, and we compute their depth dependence and frequencies to leading order. For quasi-adiabatic stratification the sectoral modes with no radial nodes are the only modes which are almost toroidal and the depth dependence of the corresponding horizontal motion scales as $r^m$. For all r modes except the zero radial order sectoral ones, non-adiabatic stratification plays a crucial role in the radial force balance.}
   {The lack of quasi-toroidal solutions when stratification is close to neutral, except for the sectoral modes without nodes in radius, follows from the statement that the system needs to be in both horizontal and radial force balance. In the absence of super- or subadiabatic stratification and viscosity, both the horizontal and radial force balances independently determine the pressure perturbation. The only quasi-toroidal cases in which the two determinations of the pressure perturbation are consistent are the special cases where $\ell=m$, and the horizontal displacement scales with $r^m$.}
   \maketitle
%________________________________________________________________
\section{Introduction}
Rossby waves, large-scale waves of radial vorticity with retrograde phase speed, have recently
been discovered from measurements of horizontal surface and near-surface solar flows
\citepads[][and confirmed by \citeauthor{2019A&A...626A...3L} \citeyear{2019A&A...626A...3L}]{2018NatAs...2..568L}. The clearly observed waves have frequencies
near those of sectoral traditional Rossby waves in a uniformly rotating fluid system \citepads[e.~g.][]{1964RSPSA.279..446L}, corresponding to sectoral spherical harmonics of azimuthal order $3 \leq m \leq 15$. \citeauthor{2018NatAs...2..568L} found that the amplitudes of these Rossby waves do not depend strongly on depth down to 21 Mm below the photosphere, but could not further characterise the radial dependence of the eigenfunctions. Assuming the motion is incompressible, they argued that viscous damping is the reason why they observe only sectoral Rossby modes in the Sun. Here we are interested in understanding, for a more realistic stellar stratification, why only the sectoral modes are seen in the observations and also, in modelling the radial structure of the observed modes.

We restrict our attention to the Rossby waves discussed in \citeauthor{2018NatAs...2..568L} which have a dispersion relation close to that of traditional (non-magnetic) Rossby waves. We do not discuss magnetic Rossby waves \citep[e.g.][]{2010ApJ...709..749Z, McIntosh, 2018ApJ...862..159D}.

The restoring force for traditional Rossby waves is the Coriolis force. Rossby waves have been studied extensively in the geophysical context \cite[see, for example, the textbook by][]{vallis_2006}, with special interest in their horizontal motion and with applications to the Earth's atmosphere and oceans \citep{Rossby,1978AnRFM..10..159D}, but also the atmospheres of Jupiter and Venus \citep[e.g.][]{1990Icar...83..282A,Venus,2019JGRE..124.1143N}.

In the stellar context, waves analogous to planetary Rossby waves are known as r modes \citepads[or quasi-toroidal modes, see e.g.][]{1978MNRAS.182..423P,1989nos..book.....U}. They have been considered in the photosphere of the Sun \citep[starting with the speculative work of][]{1966MNRAS.131..407P}, as well as near the base of the convection zone \citep[see the series of papers starting with][]{1969SoPh....8..316G} where the stratification is assumed to be subadiabatic. \citetads{1986SoPh..105....1W} studied the properties of the r modes in the convective zone of the Sun, but did not predict any restriction on the existence of non-sectoral modes for uniform rotation.

In general, it can be shown that for low-frequency nonradial oscillations of a rotating star, the spheroidal components associated to the spherical harmonics of degree $\ell$ couple with the toroidal components of adjacent degrees $\ell\pm1$. Furthermore, these toroidal components $\ell\pm1$ themselves couple with the spheroidal components associated with $\ell$ and $\ell \pm 2$. Thus, without significant simplifications, a nonradial oscillation mode in a rotating star is given by an infinite sum of terms proportional to spherical harmonics with different degrees $\ell$ for a given azimuthal index $m$ \citepads{1966AnAp...29..313Z,1978A&A....70..597B}. In numerical analysis, a truncation of the series is inevitable. Previous works have opted for drastic truncation, retaining only the first two terms \citepads[see e.g.][]{1987MNRAS.224..513L}, which may affect the results significantly.

Two approaches have been considered in order to remove this difficulty. One of these is the so-called traditional approximation, in which the horizontal component of the rotation vector is neglected \citepads{1989MNRAS.237..875L,1997ApJ...491..839L}. Then the Coriolis force associated with radial motion and the radial component of the Coriolis force associated with horizontal motion are both neglected. Alternatively, the solution can be sought using asymptotic expansion relative to a small parameter proportional to the rotation frequency \citepads{Provost, Smeyers}. The former approximation is valid locally in regions of the star where both the rotation frequency and the pulsation angular frequency in the corotating frame are significantly smaller than the Brunt-V{\"a}is{\"a}l{\"a} frequency \citepads{1989MNRAS.237..875L, 2003MNRAS.340.1020T}, whereas the latter only requires slow stellar rotation to be valid.

In this paper, we are interested in the low-degree modes observed in a slowly rotating star, so we will consider the framework of \citetads[][hereafter \citetalias{Provost}]{Provost}, concentrating on the cases where the stratification is close to, but not exactly, adiabatic. Our results are not merely an addition to those of \citetalias{Provost}, but they provide better insight into the nature of r modes, and arguably amend those of \citetalias{Provost}. Assuming that the stellar interior is inviscid and the motions are adiabatic, we show that the $\ell=m$ mode with no radial nodes is the only almost toroidal Rossby mode which can be present, for uniform rotation. The corresponding eigenfunctions scale as $r^m$.

\section{Digest of Provost et al. analysis}
This papers follows the formalism developed in \citetalias{Provost}, with minor changes in the notation. In this section, we summarise the main points of their method that we think are important for understanding our results. \citetalias{Provost} considers a rotating star, in a co-rotating reference frame of basis vectors ($\hat{\vec{e}}_r, \hat{\vec{e}}_\theta, \hat{\vec{e}}_\phi$), with the origin at the star's centre of mass and spherical coordinates ($r,\vartheta,\varphi$), where $r$ is the radial distance to the origin, $\vartheta$ is the polar angle, $\varphi$ is the azimuthal angle. Rotation is assumed to be uniform, with angular frequency $\vec{\Omega}_\star = \Omega_\star \hat{\vec{e}}_z $ parallel to $\hat{\vec{e}}_z= \cos \vartheta\, \hat{\vec{e}}_r - \sin \vartheta\, \hat{\vec{e}}_\theta$. The densities are in units of $\rho_\star = M_\star /R_\star^3$, the pressures are in units of $p_\star = GM_\star^2/R_\star^4$, the times are in units of $\tau_\star = (GM_\star/R_\star^3)^{-1/2}$ and the lengths are in units of $R_\star$, where $R_\star$ and $M_\star$ are the radius and mass of the star and $G$ the universal gravitational constant. The dimensionless angular frequency of the star is then denoted as
\begin{equation}
\varepsilon = \Omega_\star \tau_\star.
\end{equation}
For the Sun\footnote{The $\star$ subscript is replaced by the $\odot$ symbol to denote solar quantities throughout the paper.}, $\varepsilon = \Omega_\odot \tau_\odot = 4.5 \times 10^{-3}$. In a uniformly rotating star, the isobaric and isopsynic surfaces coincide with the level surfaces of constant total potential (gravitational and centrifugal). This is know as the Poincar{\'e}-Wavre theorem and holds whatever the equation of state of the gas. For slow rotators ($\varepsilon \ll 1$), those surfaces can be expressed through a distortion term of the order of $\varepsilon^2$ and a function $\alpha$ determined by the internal structure. These surface levels can be used to implicitly define a set of curvilinear coordinates $(x,\theta,\phi)$,
\begin{align}
r&= x (1-\varepsilon^2 \alpha(x) \cos^2 \theta) +O(\varepsilon^3),\label{rtransform}\\
\vartheta&=\theta + O(\varepsilon^3),\\
\varphi&=\phi + O(\varepsilon^3),
\end{align}
where the new coordinate $x$ is constant on surfaces of constant density and pressure. The surface of the star is an isobaric surface $x = $ constant and we choose the normalisation so that $x = 1$ at the surface. With respect to the new variable $x$, the equilibrium pressure $p(x)$ and density $\rho(x)$ are then independent of $\theta$ and $\phi$. The oscillations are treated as a small perturbation around this static equilibrium state. To study linear modes of oscillations, the temporal and longitudinal structure of all perturbed quantities and the displacement $\vec{\xi}$ are assumed to be proportional to\footnote{We follow the same sign conventions as in \citetads{2018NatAs...2..568L}, this means that our frequencies have the opposite sign as those of \citetalias{Provost}.}
\begin{align}
e^{i(m\phi - \sigma t)}.\label{signconv}   
\end{align}
The equations governing the small amplitude, periodic, adiabatic oscillations of a uniformly rotating star are obtained by writing the linearised equations for the conservation of angular momentum, mass, and energy:
\begin{align}
&-\sigma^2 \vec{\xi}- 2  i \sigma \varepsilon\, \hat{\vec{e}}_z \times  \vec{\xi} -\frac{ \rho'}{\rho^2} \nabla p + \frac{1}{\rho}\nabla p'=0, \label{mom0}\\
&\rho'+ \nabla \cdot (\rho  \vec{\xi}) = 0,\label{cont0}\\
&p' + \vec{\xi}\cdot \nabla p = \gamma \frac{p}{\rho} \left(\rho'  + \vec{\xi}\cdot \nabla \rho\right),\label{adiab0}
\end{align}
where $p'$ and $\rho'$ are the Eulerian perturbations of pressure and density, and $\gamma = \left(\difp{\ln p}{ \ln \rho}\right)_{\rm ad}$ is the first adiabatic exponent. Here we have neglected the perturbation of the gravitational potential (Cowling's approximation) and have assumed the flows are adiabatic and the viscosity is negligible. Cowling's approximation was shown to be justified in most cases by \citetalias{Provost}.

\citetads{1982ApJ...256..717S} argues that the adiabatic assumption is justified in stellar radiative zones except near the boundaries. The assumption that the flow is inviscid is also probably justifiable in radiative zones, however in stellar convection zones non-adiabatic mixing by turbulent convective motions and turbulent viscosity are likely to be relevant on timescales shorter than or comparable to the rotation period of the star. Notwithstanding, we chose to follow \citetalias{Provost} in considering the inviscid and adiabatic case because this reveals interesting basic physics. In addition, we expect that modes which rely for their existence on viscosity or non-adiabatic processes will decay faster than modes which exist in the adiabatic, inviscid case. Equations \eqref{mom0}-\eqref{adiab0} form a system of partial differential equations that requires appropriate boundary conditions to constitute a well posed boundary value problem. We will consider the boundary conditions at the centre and the surface of the rotating star. At the centre, the displacement must remain finite. The conservation of momentum across the non-spherical surface of the star requires that the Lagrangian pressure perturbation $\delta p =0$ at the surface.

\subsection{Series expansion in terms of \texorpdfstring{$\varepsilon$}{e}}
\citetalias{Provost} considers solutions for $\vec{\xi} = (\xi_r, \xi_\theta, \xi_\phi) , p', \rho', \sigma$, that are solutions of system Eqs.~\eqref{mom0} - \eqref{adiab0} for cases where the rotation rate is small. This motivates an expansion of the form 
\begin{align}
&\xi_r(\varepsilon)  \simeq \sum_{j=0}^4 \varepsilon^j \xi^{(j)}_r &  \xi_\theta(\varepsilon)  \simeq \sum_{j=0}^4 \varepsilon^j \xi^{(j)}_\theta  && \xi_\phi(\varepsilon)  \simeq \sum_{j=0}^4 \varepsilon^j \xi^{(j)}_\phi &  \nonumber\\
&\sigma(\varepsilon) \simeq \sum_{j=0}^4 \varepsilon^j \sigma_j & p'(\varepsilon)  \simeq \sum_{j=0}^4 \varepsilon^j p'^{(j)} & & \rho'(\varepsilon)  \simeq \sum_{j=0}^4 \varepsilon^j \rho'^{(j)}. \label{singexp}
\end{align}

Our methodology is identical to \citetalias{Provost} but our notation is slightly different. \citetalias{Provost} use the indices $j=0,1,...$ only for non-null terms in the final expansion: for example, in their notation $\sigma=\Omega \sigma^{\mathrm{Provost}}_0+ \Omega^3 (\sigma_0 \sigma_1)^{\mathrm{Provost}}$ (their equation 2). In our notation we index all terms. We thus write $\sigma=\sigma_0+ \varepsilon \sigma_1 +\varepsilon^2 \sigma_2+ \varepsilon^3 \sigma_3+ ...$. Our notation is closer to that of \citetads{Smeyers}. As is implicit in the notation of \citetalias{Provost}, symmetry arguments lead to the conclusion that the coefficients of the even powers of $\varepsilon$ in the expansion of $\sigma$ must be 0, so this difference with respect to \citetalias{Provost} is purely one of notation. Similarly an inspection of Eqs.~\eqref{mom0} - \eqref{adiab0} reveals that the even and odd coefficients in the expansion of $\xi_{\theta}, \xi_{\phi}$, $\xi_{r}$, $p'$ and $\rho'$ decouple. The solutions for the odd terms in the expansion of these quantities are trivial if we know the solution for the even powers (they correspond to the solutions for the expansion keeping only the even terms where all quantities except $\sigma$ are multiplied by $\varepsilon$).

\citetalias{Provost} is concerned with the generalisation of modes which are purely toroidal for non-rotating stars. The displacement vector of these toroidal modes satisfies $\nabla \cdot \vec{\xi} =0$ and $\xi_r =0$. They also have $\sigma^2=0$. In the case where rotation is present, these toroidal modes become non-trivial and develop characteristics similar to Rossby waves in the Earth's atmosphere and oceans. They are often referred to as r modes, after the seminal work of \citetads{1978MNRAS.182..423P}, or quasi-toroidal mode according to the nomenclature of \citetalias{Provost}. They are called quasi-toroidal because, to zeroth-order, they have the same properties as the toroidal modes ($\nabla \cdot \vec{\xi}^{(0)} =0,\, \xi_r^{(0)} =0 \text{ and } \sigma_0=0$). To summarise, the  quasi-toroidal modes are non-radial modes of low-frequency, whose radial displacement is small compared to their horizontal motion.

\citetalias{Provost} performs the equivalent of a fourth-order expansion in terms of the small parameter $\varepsilon$ in Eqs.~\eqref{mom0} - \eqref{adiab0}. At the zeroth order, the system \eqref{mom0} - \eqref{adiab0} only retains zeroth-order quantities and reduces to: \\
$\sigma_0=0 \text{ and }\xi_r^{(0)} = 0 \text{ if and only if }  p'^{(0)}  =  0, \rho'^{(0)}  =  0$, and
\begin{align}
 \difp{}{\theta}(\sin \theta \xi_\theta^{(0)})+ im \xi_\phi^{(0)} =  0, \label{vertmom0}
\end{align}
which are the non-rotating toroidal modes.  At higher-order, we have to consider the value of the Ledoux discriminant
\begin{align}
A =  \frac{1}{\rho}\difd{\rho}{x} - \left(\frac{1}{\gamma p}\right) \difd{p}{x},
\end{align}
which is a measure of convective instability as will be discussed further in Sec.~\ref{sec:res}. The case $A=0$ will be discussed later. If $A \neq 0$, the second-order approximation of the conservation of momentum ($\hat{\vec{e}}_r, \hat{\vec{e}}_\theta, \text{ and } \hat{\vec{e}}_\phi$ components) yields
\begin{align}
 &- 2  i  \sigma_1 \sin \theta \xi^{(0)}_\phi =\frac{p^{1/\gamma} }{\rho} \difp{}{x}\left(\frac{ p'^{(2)}}{p^{1/\gamma}} \right) \nonumber\\
&\qquad \qquad \qquad \quad - A g \left( \xi^{(2)}_r  -  2 \alpha \cos \theta \sin \theta \xi^{(0)}_\theta \right), \label{rmom1}\\
&\sigma_1^2 \xi^{(0)}_\theta - 2  i  \cos \theta  \sigma_1 \xi^{(0)}_\phi =\frac{1}{\rho x}  \difp{p'^{(2)}  }{\theta}, \label{vertmom1}\\
&\sigma_1^2 \xi^{(0)}_\phi + 2  i \cos \theta \sigma_1 \xi^{(0)}_\theta  = \frac{im p'^{(2)} }{\rho x \sin \theta },\label{vertmom2}
\end{align}
where \begin{align}
g = -\frac{1}{\rho}\difp{p}{x},
\end{align}
is the unperturbed gravity. The radial part of the conservation of momentum Eq.~\eqref{rmom1} involves the component of the displacement that is normal to isopotential surfaces, $\xi_x^{(2)} = \xi_r^{(2)} - 2 \alpha \cos \theta \sin \theta \xi_\theta^{(0)}$. Equation~\eqref{rmom1} is \citetalias{Provost}'s (7a). Combining our Eqs.~\eqref{vertmom1} and \eqref{vertmom2} gives their (6a), which only involves zeroth-order quantities. 

Equations~\eqref{vertmom1}, \eqref{vertmom2} and \eqref{vertmom0} express the conservation of total vertical angular momentum and of mass, to zero order. By elimination of $p'^{(2)}$, they form a classical Legendre equation. This allows us to find exact solutions for $\sigma_1$ and the angular dependence of $\xi_\theta^{(0)}$ and $\xi_\phi^{(0)}$ in terms of $\ell, m,$ and the associated Legendre polynomials. The radial dependence  of the eigenfunctions is not determined at this order. The solutions are  Rossby waves on surfaces of constant $x$, as in Eq.~(9a) of \citetalias{Provost}:
\begin{align}
\sigma_1 &= - \frac{2m}{\ell(\ell+1)}, \label{sigma1}\\
\xi_\theta^{(0)} &= im C^{(0)}_{\ell,m}(x) \frac{P_\ell^m(\cos \theta)}{\sin \theta}, \label{xitheta0}\\
\xi_\phi^{(0)} &= - C^{(0)}_{\ell,m}(x) \difd{}{\theta} P_\ell^m(\cos \theta). \label{xiphi0}
\end{align}
Here we are interested in retrograde Rossby waves so we are only going to consider $m>0$. At the fourth-order level of the approximation, the $\theta$ and $\phi$ components of the conservation of momentum 
can be combined  to eliminate the pressure perturbation terms $p'^{(4)}$ and $p'^{(2)}$, yielding Eq. (7c) from \citetalias{Provost}. Finally, the system of equations is closed by taking the second-order expansion of the continuity equation, yielding \citeauthor{Provost}'s Eq.~(7d).

After some manipulation, the aforementioned closed system of equations can be reduced to a single differential equation for the amplitude $C^{(0)}_{\ell,m}(x)$ of the horizontal displacement, Eq.~(11) of \citetalias{Provost}, 
\begin{align}
&\difd{^2}{x^2}C^{(0)}_{\ell,m}(x) + \frac{A g}{\rho x^4} \left[\difd{}{x} \left(\frac{\rho x^4}{A g}\right) \right]\difd{}{x}C^{(0)}_{\ell,m}(x)\nonumber\\
&+ \left[\frac{Ag}{x^4}\difd{}{x}\left(\frac{x^4}{g}\right) + \frac{\lambda_2}{x} \left( \frac{Ag\rho}{x^2}\difd{}{x}\left(\frac{x^2}{Ag\rho}\right) - \frac{2 \rho g}{\gamma p}\right) \right. \nonumber\\
&\left.-\frac{Ag}{x^2}\left(\lambda_3 \alpha + \lambda_4 x \difd{\alpha}{x}\right) - \frac{\ell(\ell+1)}{x^2} + \lambda_1  \frac{Ag}{x^2} \frac{\sigma_3}{\sigma_1} \right] C^{(0)}_{\ell,m}(x) =0, \label{eq11}
\end{align}
where $\lambda_1,\lambda_2,\lambda_3$ and $\lambda_4$ are functions of $m$ and $\ell$ and are defined in \citetalias{Provost}. When  $A\neq0$, for $x\in (0,1)$ and with appropriate boundary conditions, Eq.\eqref{eq11} is a Sturm–Liouville eigenvalue problem.

\subsection{Boundary condition at the centre}
Near the centre $x\sim 0$, the quantities that appear in Eq.~\eqref{eq11} are in order of magnitudes as follows
\begin{align}
    \rho &\sim \rho_c,\\
    p &\sim p_c,\\
    g &\sim g_c x, \\
    A &\sim A_c x, \\
    \alpha &\sim 0,
\end{align}
where the subscript $c$ denotes values at the centre. Therefore, by asymptotic analysis of Eq.~\eqref{eq11} in the vicinity of $x=0$, we have
\begin{align}
  &\difd{^2}{x^2}C^{(0)}_{\ell,m}(x) + \frac{2}{x} \difd{}{x}C^{(0)}_{\ell,m}(x) - \frac{\ell(\ell+1)}{x^2} C^{(0)}_{\ell,m}(x) =0.\label{eq11in0}  
\end{align}
Using the Frobenius method \citepads{1999amms.book.....B}, we look for a solution of the form $C^{(0)}_{\ell,m}(x) = \sum_{j=0}^\infty a_j x^{j+\beta}$ in the vicinity of $x=0$. According to \eqref{eq11in0}, the coefficient of the lowest power of $x$ must satisfy
\begin{align}
\beta(\beta+1)-\ell(\ell+1)=0,  
\end{align}
which has solutions $\beta=\ell$ and $\beta =-\ell-1$. The only solution that ensures the regularity of $C^{(0)}_{\ell,m}$ is $C^{(0)}_{\ell,m}(x)\sim x^{\ell}$ in the vicinity of $x=0$. Hence we use the following boundary condition
\begin{align}
&\lim_{x\rightarrow 0}\left(\difd{C^{(0)}_{\ell,m}}{x} - \frac{\ell}{x} C^{(0)}_{\ell,m}\right)=0 \label{BC1}
\end{align}
and
\begin{align}
\lim_{x\rightarrow 0} C^{(0)}_{\ell,m}(x)=0. \label{BC2}
\end{align}

\subsection{Boundary condition at the surface}
At the free surface the Lagrangian pressure perturbation 
\begin{align}
 \label{deltap}
    \delta p =& \frac{4 i m (\ell+m) \rho}{\ell^2(2\ell+1) A} \left( x \difd{C^{(0)}_{\ell,m}}{x} +(\ell +1) C^{(0)}_{\ell,m}\right) P_{\ell-1}^m(\cos \theta) \nonumber\\
    &+ \frac{4 i m (\ell-m+1) \rho}{(\ell+1)^2 (2\ell+1) A} \left( x \difd{C^{(0)}_{\ell,m}}{x} - \ell C^{(0)}_{\ell,m}\right) P_{l+1}^m(\cos \theta) 
\end{align}
must vanish.  This is a very stringent condition for the existence of non-sectoral Rossby waves in the inviscid case, and its implications depend on whether $\rho/A$ vanishes or not at the surface.

\subsubsection{Boundary conditions for a star where the surface has \texorpdfstring{$\rho/A=0$}{vanishing density}}\label{fullpoly}
As a model of a star \citetalias{Provost} considers a complete polytrope\footnote{Polytropes will be described in more details in Sec.~\ref{sec:res}.} characterised by a polytropic index $n$. The stellar surface for such models is defined by $\rho=p=0$.

In general, for a complete polytrope or otherwise, when $\rho/A$ vanishes at the surface, $\delta p=0$ is met at the surface as long as $C^{(0)}_{\ell,m}$ and its derivative are regular. This is the boundary condition that was used in \citetalias{Provost}.

For the particular case of a polytropic stellar model,
the behaviour of the quantities that appear in \eqref{eq11} can be approximated by Taylor expansion of $\hat{\theta}$, the solution to the Lane-Emden equation, near $x=1$ as follows

\begin{align}
    \rho &\sim \rho_c \left(-\left.\difd{\hat{\theta}}{x}\right|_{x=1}\right)^n (1-x)^n,\\
    p &\sim p_c \left(-\left.\difd{\hat{\theta}}{x}\right|_{x=1}\right)^{n+1}(1-x)^{n+1},\\
    g &\sim - \frac{(n+1) p_c}{\rho_c}\left.\difd{\hat{\theta}}{x}\right|_{x=1}, \\
    A &\sim - \left(n-\frac{n+1}{\gamma}\right)\frac{1}{1-x}, \\
    \alpha &\sim \alpha(1),
\end{align}

By asymptotic analysis of Eq.~\eqref{eq11} in the vicinity of $x=1$, we have
\begin{align}
  &\difd{^2}{x^2}C^{(0)}_{\ell,m}(x) - \frac{n+1}{1-x} \difd{}{x}C^{(0)}_{\ell,m}(x) + \frac{Q(x)}{1-x} C^{(0)}_{\ell,m}(x) =0,\label{eq11in1}  
\end{align}
where 
\begin{align}
\label{QQ}
    Q(x) = \left( n - \frac{n+1}{\gamma } \right) \left[  \frac{ (n+1) p_c }{\rho_c} \left.\difd{\hat{\theta}}{x}\right|_{x=1} \left( \lambda_1 \frac{\sigma_3}{\sigma_1} - \lambda_3 \alpha\right) -1\right]\nonumber  \\
  + \lambda_2 \left( n-1 - \frac{2( n+1) }{\gamma}\right).
\end{align}
We remind the reader that $\lambda_1$, $\lambda_2$ and $\lambda_3$ are simple, but tediously long, rational functions of $\ell$ and $m$. They are given explicitly in \citetalias{Provost}.

Thus the condition of regularity for $C^{(0)}_{\ell,m}(x)$ at the surface is
\begin{align}\label{regsurf}
\left\{ (n+1) \difd{C^{(0)}_{\ell,m}}{x} - Q(x) C^{(0)}_{\ell,m} \right\}_S=0. 
\end{align}
Here and from now on, the notation in bracket with subscript $S$ means that we take the value at the surface.

\subsubsection{Boundary conditions when \texorpdfstring{$\rho/A \neq 0$}{density does not vanish} at the surface}\label{nonsectoral}
Truncated polytropes have been used as models of stars which include an atmosphere \citepads[e.g.]{Hendry1993, Bogdan}. In these models, the polytrope is truncated at some location, with non-zero pressure and density, that represents the stellar surface. In the limit that the density of the atmosphere is small compared to the surface density, the boundary condition corresponds to a free surface (with vanishing Lagrangian pressure perturbation at the surface). Such truncated polytropes with a free surface have previously been used in the study of  helioseismic acoustic waves \citep[e.g.][]{Bogdan}. The analysis for Rossby waves proceeds differently according to whether $\ell=m$ or not.

\paragraph{Cases where $\ell = m$.}
When $\ell= m$, the free-surface boundary yields one condition
\begin{align}
  &\left\{ \frac{\rho}{A}\left( x \difd{C^{(0)}_{m,m}}{x} - m C^{(0)}_{m,m} \right)\right\}_{S}=0,
\end{align}

In the cases where $\{\rho/A\}_S \ne 0$, the boundary condition reduces to 
\begin{align}\label{BC3}
  &\left\{  \difd{C^{(0)}_{m,m}}{x}\right\}_S = \left\{\frac{m}{x} C^{(0)}_{m,m}\right\}_{S}.
\end{align}

\paragraph{Cases where $\ell \ne m$.}
 When $\ell\neq m$, the free-surface boundary condition involve two associated Legendre polynomials $P^m_{\ell-1}$ and $P^m_{\ell+1}$, whose coefficients must both vanish and this thus yields two conditions\footnote{The same boundary conditions were already given in \citetads{Smeyers} Eqs.~(65) and (66). There is however a sign error in their Eq.~(65), which is corrected in our Eq.~\eqref{surf2}.}
\begin{align}
 &\left\{ \frac{\rho}{A}\left( x \difd{C^{(0)}_{\ell,m}}{x} - \ell C^{(0)}_{\ell,m} \right)\right\}_{S}=0, \label{surf1}\\
   &\left\{  \frac{\rho}{A}\left( x \difd{C^{(0)}_{\ell,m}}{x} + (\ell+1) C^{(0)}_{\ell,m} \right)\right\}_{S}=0. \label{surf2}
\end{align}
In general, $\rho/A$ does not vanish at the surface of stars,  Eqs.~\eqref{surf1} and \eqref{surf2} can only be met at the same time by requiring both $\difd{C^{(0)}_{\ell,m}}{x}~=~0$ and $C^{(0)}_{\ell,m}~=~0$ at the surface. These two boundary conditions, together with the boundary condition at the centre Eq.~\eqref{BC1} fully specify the problem and the only solution is then the trivial solution $C^{(0)}_{\ell,m} (x)=0$.

Consequently, for non-vanishing density at the surface, there is no non-trivial solution when $\ell \neq m$. The sectoral mode is the only quasi-toroidal mode that can satisfy the free-surface boundary condition in the inviscid case. Let us note that the non-sectoral modes found by \citetads{1986SoPh..105....1W}, using \citetalias{Provost}'s derivations applied to the Sun, are obtained by imposing $\difd{C^{(0)}_{\ell,m}}{x}~=~0$ at $x=0.999$, which is inconsistent with a free-surface boundary condition. We remark as well that the no-penetration boundary condition is $\xi_x^{(2)}=0$ with
\begin{align}
    \xi_x^{(2)} &= \frac{4 i m (\ell-m+1) }{(\ell+1)^2 (2\ell+1)Ag} \left( x \difd{C^{(0)}_{\ell,m}}{x}+ (A x - \ell) C^{(0)}_{\ell,m}\right) P_{l+1}^m(\cos \theta) \nonumber\\
    &+ \frac{4 i m (\ell+m) }{l^2 (2\ell+1)Ag} \left( x \difd{C^{(0)}_{\ell,m}}{x} +(A x + \ell +1) C^{(0)}_{\ell,m}\right) P_{\ell-1}^m(\cos \theta) ,
\end{align}
so that \citetads{1986SoPh..105....1W} appears to be also inconsistent with the no-penetration boundary condition.

\subsection{Comparison of boundary conditions of polytrope and truncated polytrope.}
It is interesting to compare the boundary conditions for the complete and truncated polytropes. For $\ell=m$ these are Eqs.~\ref{regsurf} and \ref{BC3} respectively. The quantity $Q(1)$ is given by Eq.~\ref{QQ} evaluated at the surface and depends on $m$. For the special cases where $\ell=m$, $\gamma=5/3$ and $n=3/2$, we find that $Q(1)/(n+1)=m$. The values of $Q/(n+1)$ evaluated at the surface of the full polytrope ($x=1$) are given in Table~\ref{table:1} for some of the other cases studied in this paper with $\ell=m=3$. 
In Table~1 we see that $Q/(n+1)$ is approximately equal to $m$ near $n=1.5$
(it is exactly $m$ at $n=1.5$). A consequence of this is that in the neighbourhood of $n=1.5$, the boundary condition for the truncated polytrope is equivalent to that for the complete polytrope. This indicates that the results have some robustness to the details of the model. 

\begin{table}
\caption{Value of $Q/(n+1)$ evaluated at $x=1$ for $\ell=m=3$ for different polytropic indices. The corresponding quantity involved in the free-surface boundary conditions for the truncated polytrope is $\sim m$, whatever the value of $n$.} 
\label{table:1}      
\centering                                      
\begin{tabular}{c c c c c c c c}          
\hline\hline                        
$n$ & 1 & 1.25 & 1.49 &1.51 & 2.5 & 3\\    
\hline                                   
$Q(1)/(n+1)$ & 6.37 & 4.11 & 3.03 & 2.97 & 2.77 & 3.51\\
\hline
\end{tabular}
\end{table}

For $\ell \ne m$ the situation is more complicated, and no solutions exist for the truncated polytrope. This is also the case for the complete polytrope when $n=1.5$, (see Appendix I in \citetalias {Provost}). For other values of $n$, the solution for the complete polytrope of course exists, and is as described in \citetalias{Provost}.

\section{Results}\label{sec:res}
 To solve the boundary-value problem for \eqref{eq11}, we use an implementation of a fourth-order collocation algorithm based on control of residuals, provided by the function \texttt{solve\_bvp} from the \texttt{scipy.integrate} python module. We follow \citetalias{Provost} in considering polytropes, characterised by a polytropic index $n$. In the stellar context, a polytrope is a gas spheroid in gravitational equilibrium, where the pressure is related to the density by the relation
 \begin{align}\label{polytrope}
     p = K \rho^{1+\frac{1}{n}},
 \end{align}
where $K$ is a constant of proportionality. By definition of the free surface, a complete polytrope has $p=0$ and $\rho=0$ at the surface. For all polytropes, the Ledoux discriminant $A$ is a monotonic function of $x$ and it is everywhere positive when $n<1.5$ (super-adiabatic stratification). The polytrope with index $n=1.5$ has the interesting property that then $A=0$ everywhere in the star (adiabatic stratification). Finally, for $n>1.5$, $A$ is everywhere negative (sub-adiabatic stratification. The Ledoux discriminant is the argument of the criterion for convection which develops or not whether $A>0$ or $A<0$  \citepads{1958HDP....51..353L}. So we shall also refer to the convectively unstable case when $n<1.5$, neutrally stratified case when $n=1.5$, and radiative case when $n>1.5$.

The shape function $\alpha(x)$ in Eq.~\eqref{rtransform} can be derived for the distorted polytrope \citepads{1933MNRAS..93..390C}, and is obtained by solving the Lane-Emden equation modified for rotation. We obtain the full polytrope by setting the surface $x=1$ at the first zero of the Lane-Emden function, and the truncated polytrope by setting the surface at $x=0.999$.

To test our solver, we computed the eigenvalues for $n=1$ and $n=3$ given in Tables~1 and 2 of \citetalias{Provost} to the same decimal place accuracy, for a complete polytrope using the boundary conditions \eqref{BC1}, \eqref{BC2} and \eqref{regsurf}.  We also reproduced the radial dependence of $C^{(0)}_{\ell,m}(x)$ for the same $(\ell, m, k)$ as in their figures~1, 2 and 3.

We also computed the eigen-solutions for the truncated polytropes when $\ell=m$, using the boundary conditions \eqref{BC1}, \eqref{BC2} and \eqref{BC3}. As is to be expected from Table~1, for sectoral modes, the eigen-solutions for the full polytrope and the same polytrope truncated very close to the surface are nearly identical close to $n=1.5$, and the differences even for $n=1$ and $n=3$ are small and not distinguishable in the figures that follow. Hence we only show solutions for a polytrope truncated at $x=0.999$. 

For cases when $\ell \ne m$, we refer the reader to \citetalias{Provost} for the solutions in a complete polytrope and again note that there are no such solutions satisfying the free-surface boundary conditions for truncated polytropes.

\subsection{Non-adiabatic stratification is essential to the normal force balance for all sectoral r-modes except those with no radial nodes}

%--------------------------------
\begin{figure}
   \centering
   \includegraphics[width=\columnwidth]{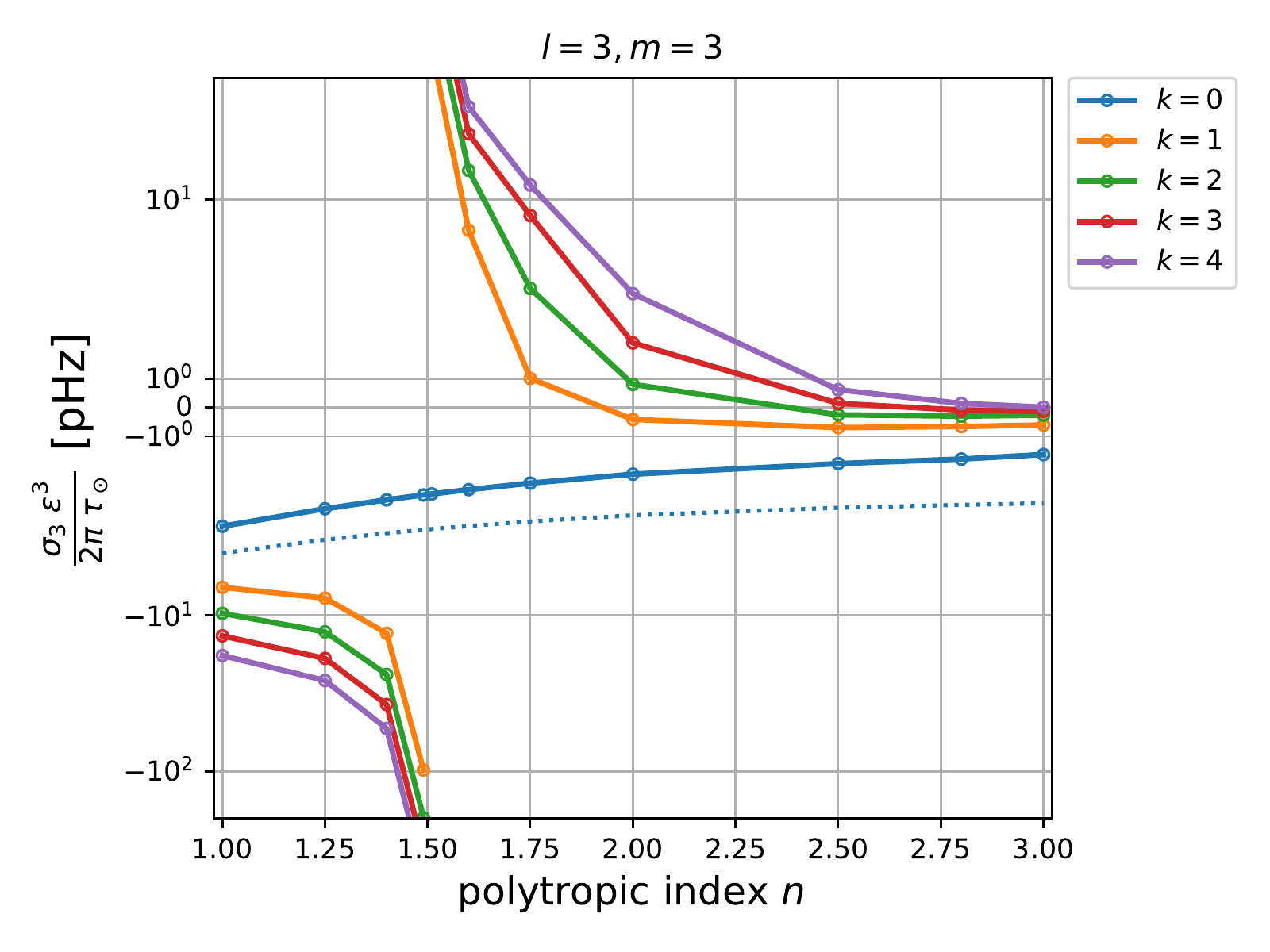}
   \caption{Third-order term in the frequency expansion $\sigma_3 \varepsilon^3$ as a function of the polytropic index $n$, for the sectoral r modes $\ell=m=3$ and several radial orders $k$. The corresponding frequency term in the incompressible case, which has only $\ell=m$, $k=0$ solutions, is given by a blue dotted line (see text). The frequency is given in pHz, using the relevant solar quantities, and is displayed in symmetric log scaling, with linear scaling between $\pm6$~pHz. The quantity $\tau_\odot$ is our unit of time and corresponds to the Sun's dynamical time scale. }
              \label{sigma3}%
\end{figure}
%-----------------------  

To get a better understanding of the effect of stratification on the r modes, we also solve the problem for $1 \leq n \leq 3$. As an example, Figure~\ref{sigma3} shows the eigenfrequencies found for $\ell=3$, and $m=3$ as a function of the polytropic index $n$ for several values of the radial order $k$. We see that for all modes, with the notable exception of the sectoral mode with $k=0$, the eigenfrequencies become increasingly large as $n$ gets close to $1.5$ (where $A=0$). Let us stress here that the derivation of Eq.~\eqref{eq11} is obtained by a singular perturbation method. Thus the asymptotic expansion \eqref{singexp} does not necessarily converge, and the solution is valid when $|\sigma_3| \ll \sigma_1/\varepsilon^2$, which means here $\varepsilon^2 |\sigma_3| /(2 \pi \tau_\odot)~\ll~10^5$~pHz for $\ell=m=3$  (see Fig.~\ref{sigma3}). 

\begin{figure*}
   \centering
    \includegraphics[width=\columnwidth]{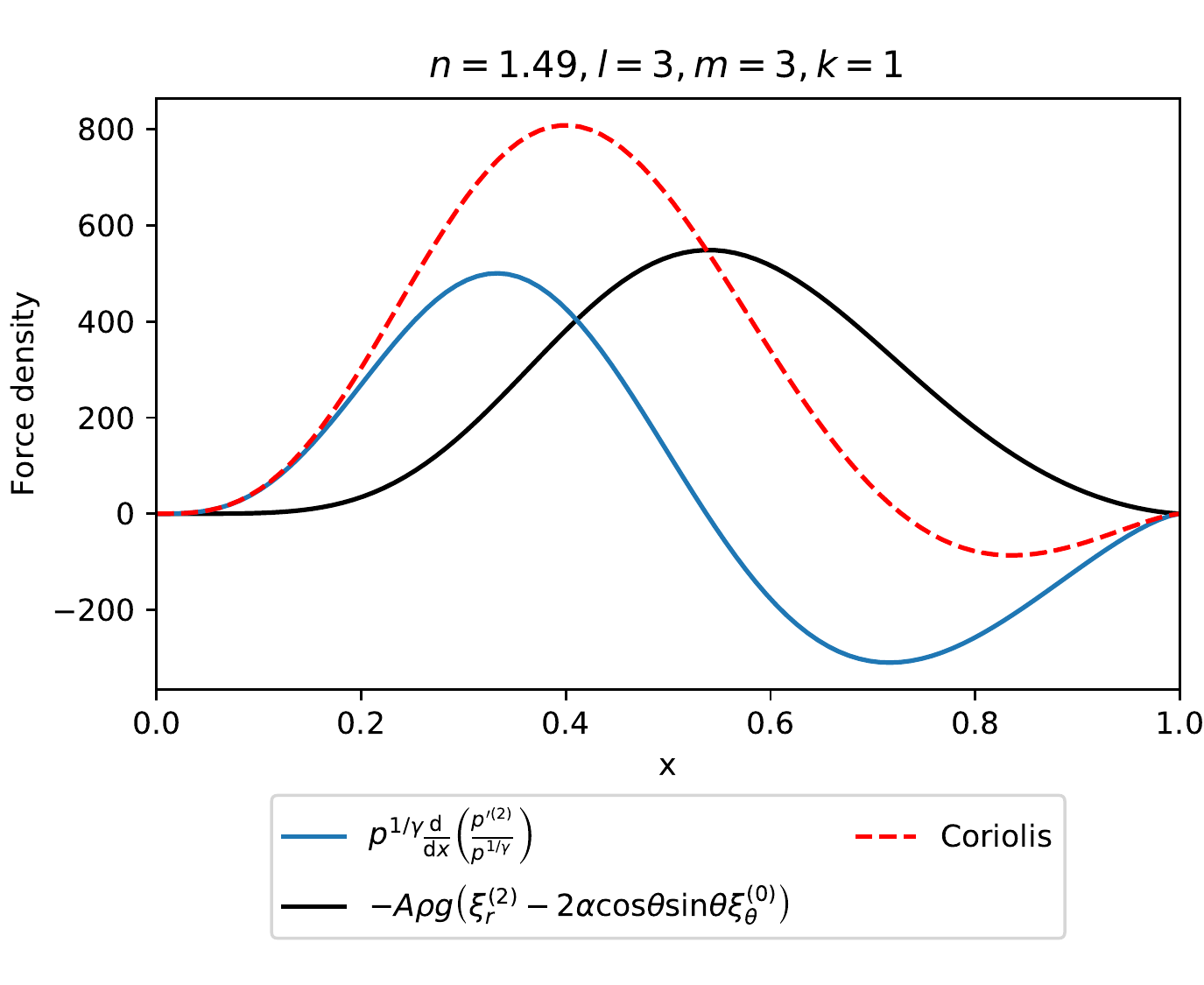}
    \includegraphics[width=\columnwidth]{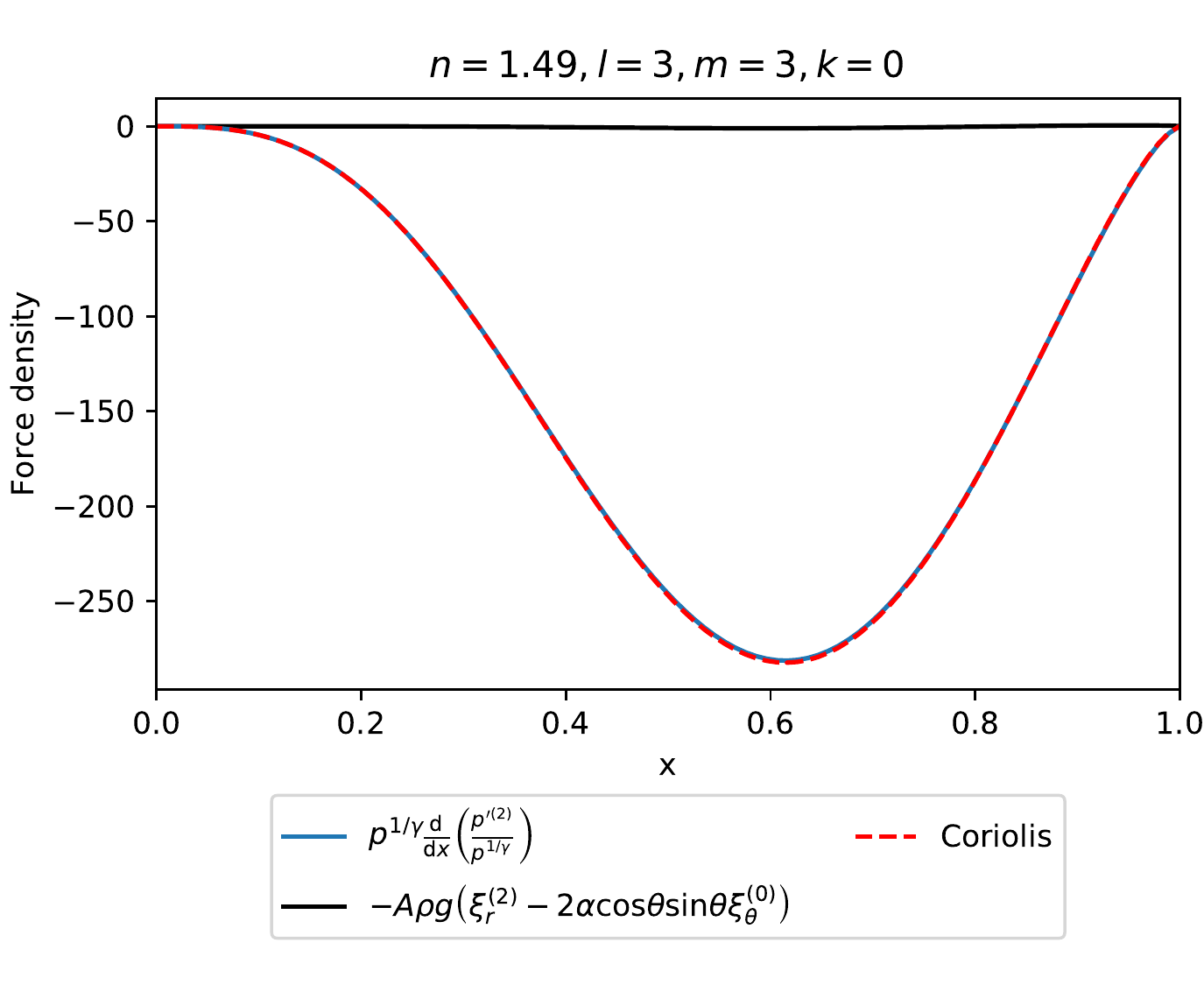}
   \caption{Terms involved in the normal force balance at $\theta=0.2$, according to Eq.\eqref{rmom1} for the slightly sub-adiabatic polytropic index $n=1.49$, in the case of the $(\ell,m,k)=(3,3,1)$ mode (left) and the (3,3,0) sectoral mode of zero radial order (right). All the quantities are dimensionless but they have been normalised consistently with Fig.~\ref{pertub331}.}
    \label{forcebal}
\end{figure*}

Solving Eq.~\eqref{eq11} allows the derivation of all the perturbed quantities of our problem.  Figure~\ref{forcebal} shows the radial force balance associated to r modes given by Eq.~\eqref{rmom1} for the polytrope $n=1.49$ and for $\ell=m=3$. In the momentum equation, the Coriolis term is balanced by a non-trivial combination of $\frac{p^{1/\gamma} }{\rho} \difp{}{x}\left(\frac{ p'^{(2)}}{p^{1/\gamma}} \right)$ and  $-A g \left( \xi^{(2)}_r  -  2 \alpha \cos \theta \sin \theta \xi^{(0)}_\theta \right)$. For the modes with $k\neq0$ (Fig.~\ref{forcebal}, left) the non-adiabatic stratification plays an essential role in the radial force balance. This turns out to be true for all the modes (i.e. also those with $\ell \neq m$) except the $\ell=m$, $k=0$ modes (this will be discussed in Section~4). For the case with $\ell=m$ and $k=0$, the radial force balance is essentially between the Coriolis term and $p^{1/\gamma} \difp{}{x}\left(\frac{ p'^{(2)}}{p^{1/\gamma}} \right)$ (Fig.~\ref{forcebal}, right) -- the term involving the non-adiabatic stratification plays essentially no role.  This is a special property of the $\ell=m$, $k=0$, r modes. 

%--------------------------------
%-----------------------------

\subsection{The depth dependence of the sectoral modes of zero radial order is \texorpdfstring{$x^m$}{a power-law} for quasi-adiabatic stratification}
%-----------------------  
\begin{figure}
    \centering
    \includegraphics[width=\columnwidth]{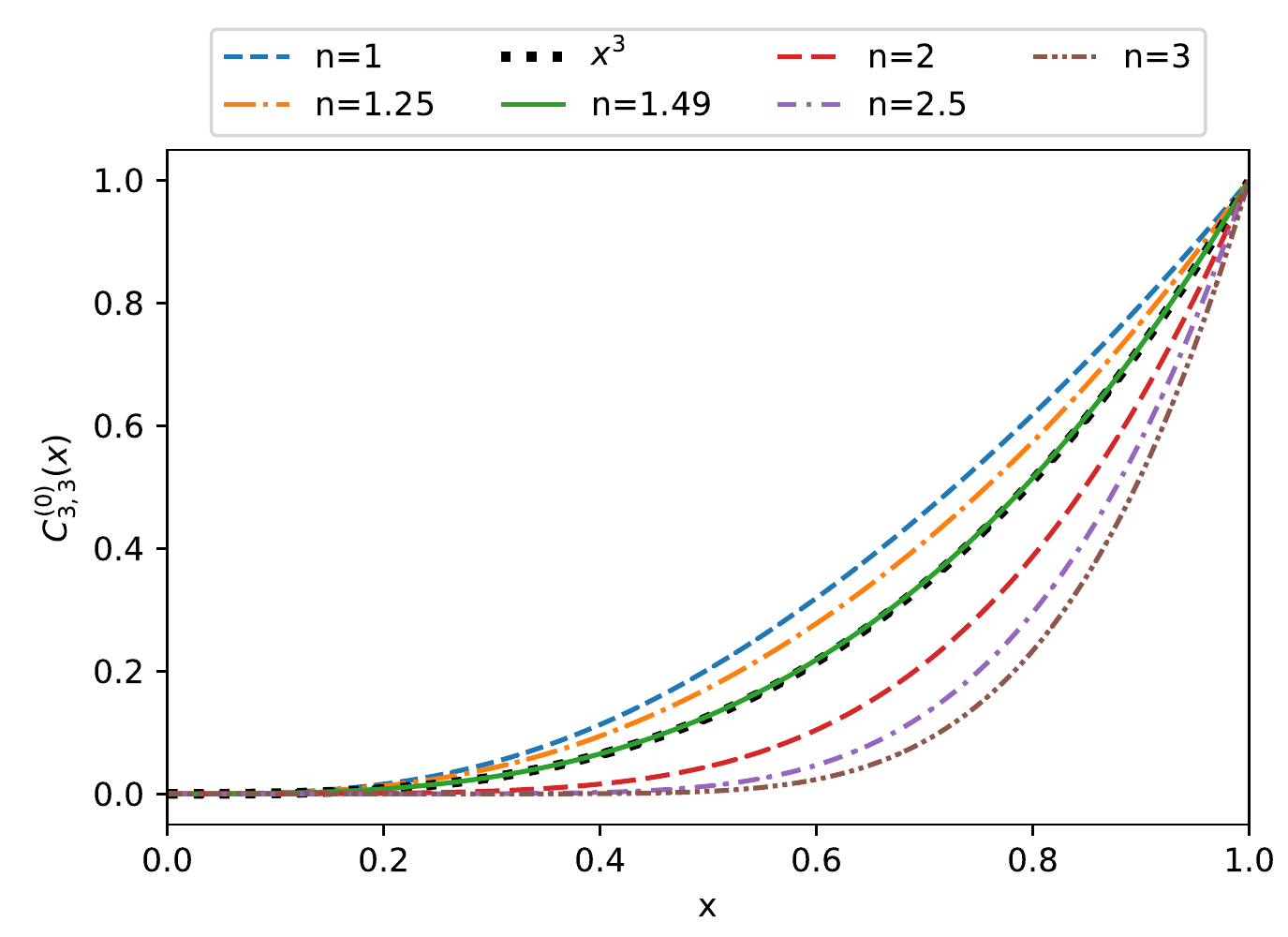}
   \caption{Solutions to Eq.~\eqref{eq11} for the sectoral modes $\ell=m=3$ of zero radial order for different polytropic indices. The functions have been normalised in order to have $C^{(0)}_{3,3}(0.999)=1$ in all cases.}
        \label{Clms}%
    \end{figure}
    
Figure~\ref{Clms} shows the radial structure $C^{(0)}_{\ell,m}(x)$ for  $(\ell, m, k)=(3,3,0)$ for different values of the polytropic index $n$. As shown by \citetalias{Provost}, when $A=0$, the only non-trivial modes are the sectoral modes with $\ell=m$. The solution must then have the form $C^{(0)}_{m,m}(x)~=~x^m$, which has no radial nodes. However, in this case there is no finite value of $\sigma_3$ that can give a finite radial displacement at the surface of the complete polytrope, since in this case the solution is divergent at the surface \citepalias{Provost}. This problem does not exist for truncated polytropes. 

In the limit of $n\rightarrow 1.5$, as we can see in Fig.~\ref{Clms}, we find that the depth dependence of the sectoral mode with $k=0$ is very close to $x^m$. Since this mode is the only one allowed to exist in the case $A=0$, we find that there is no discontinuity of solutions for this mode, and the solutions slowly depart from an $x^m$ dependence as the stratification departs from neutral.

The solution near $n=1.5$ (where $A=0$) has a radial dependence proportional to $x^m$. This is also the form of the incompressible Rossby wave \citep{1889RSPTA.180..187B,Provost}, which has no dependence on the stratification, as will be discussed in Section~\ref{sec:discuss}.

\subsection{Symmetries of the eigenfunctions about \texorpdfstring{$n=1.5$}{n=1.5} }
\begin{figure*}
    \centering
    \includegraphics[width=\textwidth]{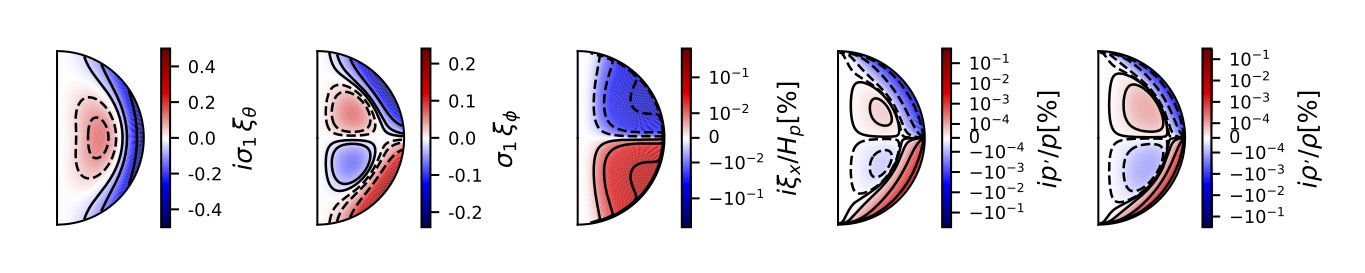}
    \includegraphics[width=\textwidth]{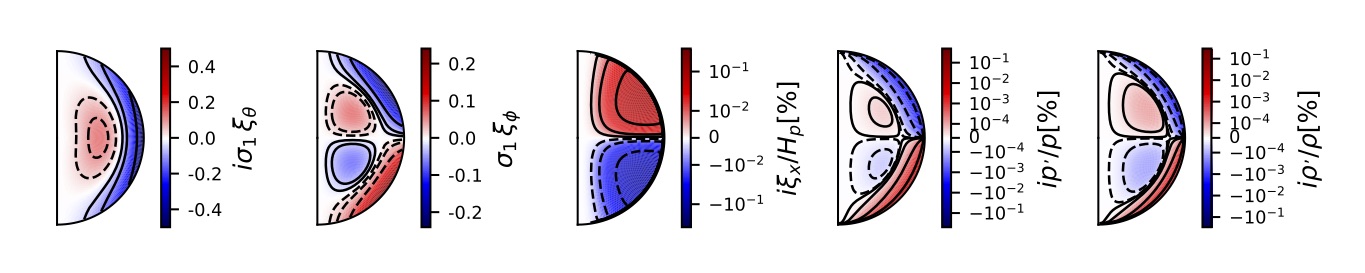}\\
       \caption{Meridional cuts $(x,\theta)$ showing the leading order terms, for solar rotation, of the latitudinal  $\sigma_1 \xi_\theta$ and azimuthal $\sigma_1 \xi_\phi$ flow velocity, the relative normal displacement $\xi_x/H_p$, the relative pressure $p'/p$ and density $\rho'/\rho$ perturbations (from left to right respectively, as labelled in the colour bar) for the $(\ell,m,k)=(3,3,1)$ mode and for $n=1.49$ (top) and $n=1.51$ (bottom). $H_p$ is the pressure height scale. The radial dependence of all the quantities has been normalised consistently so that at phase $0$, a dimensionless velocity of 1 on this scale corresponds to a velocity of $1$~m/s on the Sun.}
              \label{pertub331}%
    \end{figure*}
    
\begin{figure*}
    \centering
    \includegraphics[width=\textwidth]{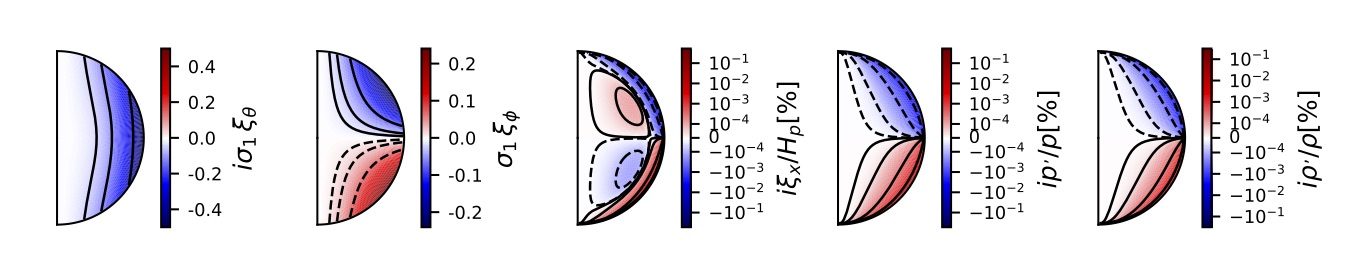}\\
    \includegraphics[width=\textwidth]{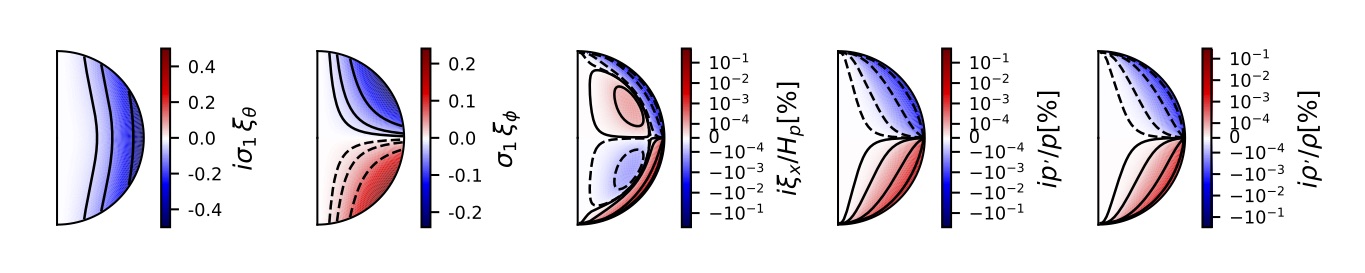}
   \caption{Same as Fig.~\ref{pertub331} but for the $(\ell,m,k)=(3,3,0)$ mode. From left to right:  latitudinal and azimuthal flow velocity, relative normal displacement, relative pressure and density perturbations for $n=1.49$ (top) and $n=1.51$ (bottom).}
              \label{pertub330}%
    \end{figure*}
    
Figs.~\ref{pertub331} and~\ref{pertub330} shows eigenfunctions for the $(3,3,1)$ and $(3,3,0)$ modes, both for $n=1.49$ and $n=1.51$. These values of $n$ were chosen because the background is convectively stable in one case and unstable in the other. Here, instead of the radial component of the displacement, we plot the more relevant component of the displacement $\xi_x^{(2)}$ that is normal to isopotential surfaces. 

The solution for the case with $k\neq0$ (Fig.~\ref{pertub331}) shows that most of the physical quantities vary only slightly across $n=1.5$ except for $\xi_x$, which flips sign. The reason for this flip in sign can be inferred from Eq.~\eqref{rmom1}, and will be discussed in Section~4. Conversely, in the case $\ell=m$ and $k=0$ (Fig.~\ref{pertub330}) all the quantities vary only slightly across $n=1.5$ and in particular the sign of $\xi_x$ does not change.  

This difference results from the fact that the radial force balance in the cases with $\ell \ne m$ or $k\ne 0$ the super-- or sub--adiabaticity plays an essential role and $\xi_x$ changes sign with $A$. By way of contrast, the sectoral mode of zero radial order does not depend essentially on the adiabaticity and $\xi_x$ varies smoothly with $n$.

\section{Discussion}\label{sec:discuss}

The solutions with $\ell=m$ and $k=0$ are qualitatively different from all other solutions.
We showed the mode with $(\ell, m, k)=(3,3,0)$ as an example. In the neighbourhood of $n=1.5$ (corresponding to $A \sim 0$), all values of $\ell$, $m$, and $k$ admit solutions in a complete distorted polytrope, as discussed above. Only the sectoral modes ($\ell=m$) would be admissible if the density at the surface did not vanish. For the cases with $\ell=m$ and $k=0$, the term $A \xi_r^{(2)}$ remains small and the solution is not sensitive to $A$; the solution does not depend on non-adiabatic stratification. All other solutions are baroclinic modes and the term  $A \xi_r^{(2)}$  is approximately constant in the neighbourhood of $n=1.5$, with $|\xi_r|$  being arbitrarily large in the limit $n\rightarrow 1.5$ ($A \rightarrow 0$) and flipping sign at $n=1.5$. Similar reasoning explains the behaviour shown in Fig.~\ref{sigma3} for the eigenfrequencies.

The radial displacements associated with solutions where the non-adiabatic stratification is essential (i.e., where the contribution to the radial force balance from $A g \xi_r$ is substantial),  go to infinity as $n$ approaches 1.5 (where $A\rightarrow 0$). These solutions cease to be quasi-toroidal. Hence the only solutions which are valid\footnote{Note that even this solution has problems at $A=0$ if the surface of the star is assumed to have zero pressure and temperature \citepalias[see ][]{Provost}.} near $n=1.5$ are those with $\ell=m$ and $k=0$.

The lack of quasi-toroidal solutions as $n$ approaches 1.5, except when $\ell=m$ and $k=0$, is not a consequence of the chosen expansion. It follows from the statement that the system needs to be in both horizontal and radial force balance. In the absence of super- or subadiabatic stratification and viscosity, both the horizontal and radial force balances independently determine the pressure perturbation. The only case in which the two determinations of the pressure perturbation are consistent and quasi-toroidal are the special cases where $\ell=m$, $k=0$ and the horizontal displacement scales with $x^m$. It is here that we make contact with the example Rossby waves in an incompressible, unstratified spherical shell discussed by \citetads{2018NatAs...2..568L}.

The existence of non-trivial solutions only in the $\ell=m$ case with an $r^m$ radial dependence is also that found by  \citetalias{Provost} in the incompressible case with arbitrary stratification. This is understood by considering that the horizontal force balance sets the horizontal structure of the pressure perturbation independently of $A$ or incompressibilitity. In the incompressible case (or in the case with $A=0$) and for quasi-toroidal motion, the pressure perturbation must alone balance the radial component of the Coriolis force, and this is only possible in the case $\ell=m$, and results in an $r^m$ dependence, and has no dependence on stratification.

The expression of the third-order term $\sigma_3$ of the frequency expansion for the $\ell=m$ mode in the incompressible case for non-spherical shapes have been obtained by \citetads{1889RSPTA.180..187B} and \citetalias{Provost},
\begin{align}
\sigma_3 =  \frac{8}{(m+1)^4} - 4 \alpha_{x=1} \frac{m}{(m+1)^2}. \label{siginc}
\end{align}
Fig.~\ref{sigma3} shows these values of $\sigma_3$ for $m=3$ as a function of $n$ as a blue dotted line\footnote{In Eq.~\eqref{siginc} only the $\alpha$ coefficient is a function of $n$, because the shape distortion depends on density distribution.} ; they are the same order of magnitude as the eigenfrequency associated to the sectoral mode of zero radial order. This is to be expected since here, we are seeking solutions that are quasi-toroidal, which means that the flow is divergence free to zeroth-order. 

All these solutions are approximately the same, although the problem is formulated differently in each case (compressible vs incompressible, polytrope vs neutrally stable stratification). In the incompressible case or the neutrally stratified case, the radial force balance is independent of the normal displacement. In this case, the problem admits only solutions that have $\ell=m$ and $k=0$ and an $x^m$ dependence. For the quasi-toroidal modes, the solutions are again approximately the same because in this case also the toroidal components of the motion are the same (spherical harmonics characterised by $\ell, m$). This in turns determines the pressure perturbation, which then must balance the radial component of the Coriolis force if the motions are to remain quasi-toroidal. This only happens for $\ell=m$ and $k=0$. For $\ell \neq m$ or $k\neq0$, the modes will develop substantial radial velocity whenever the stratification is close to, but not strictly, neutral.

\subsection{Non-spherical geometry is important away from \texorpdfstring{$n=1.5$}{n=1.5}}
\begin{figure}
    \centering
    \includegraphics[width=0.98\columnwidth]{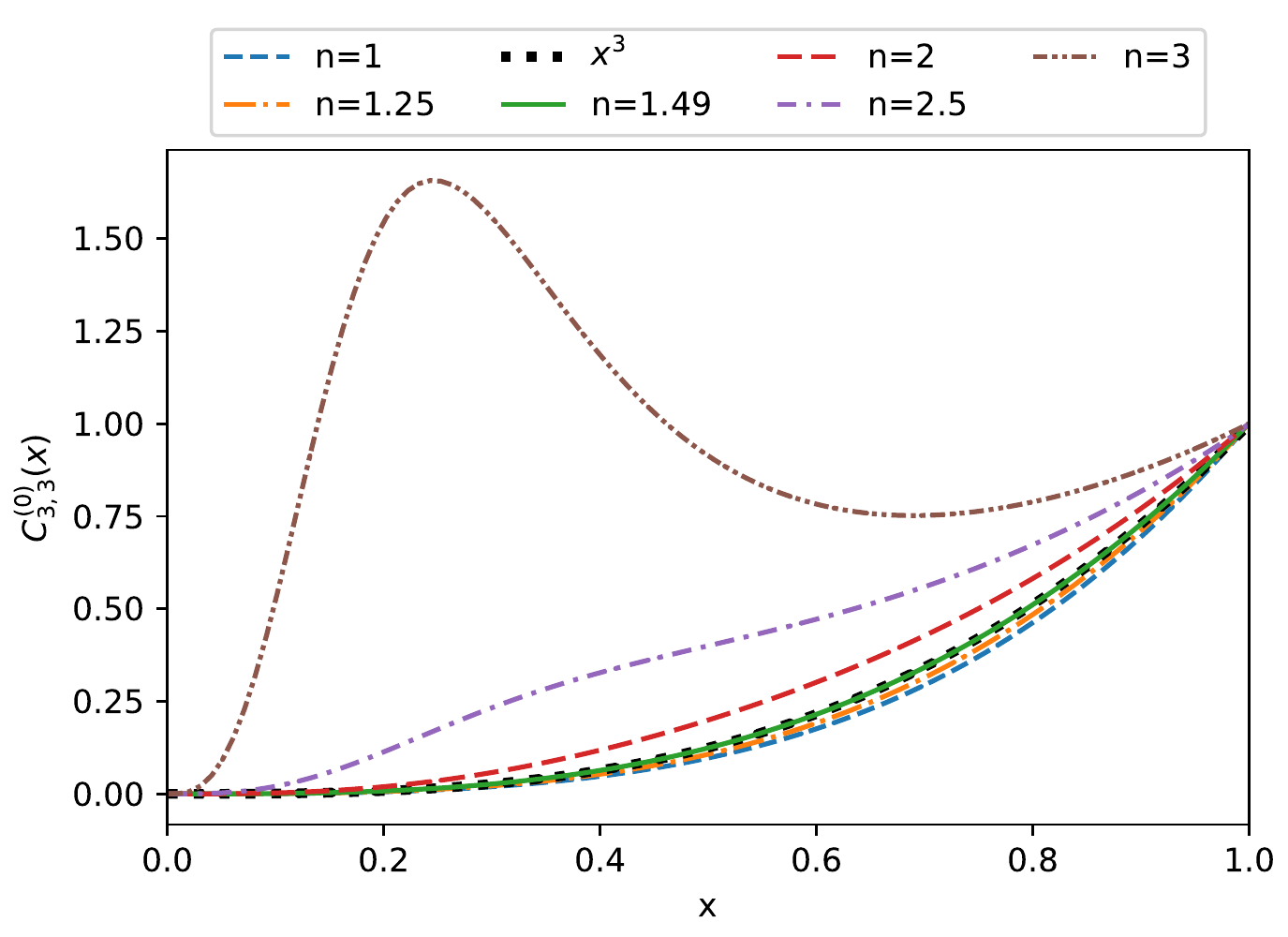}
   \caption{Same as Fig~\ref{Clms} for the $(\ell,m,k)=(3,3,0)$ mode except with spherical geometry assumed by setting $\alpha=0$. The radial structure of the eigenfunctions is affected by omission of the shape distortion when $n\neq1.5$.}
        \label{Clms_2}%
    \end{figure}

We also found the eigenvalues and eigenfunctions for the case where we artificially set $\alpha=0$, which then treats the star as a perfect sphere. Consistent with \citetalias{Provost}, we found the changes to $\sigma_3$ were substantial relative to $\sigma_3$ (i.e. can have the wrong sign), but since $|\sigma_3| \varepsilon^3$ is small, this amounts to a change of the order of $10$~pHz for the solar rotation rate. More important are the changes in the radial structure of the eigenfunctions, which can be seen by comparing Fig.~\ref{Clms_2} with the equivalent in Fig.~~\ref{Clms}. This would suggest that the distortion of the background star due to rotation should be included when determining the radial structure of the eigenfunctions, especially for non-adiabatic stratifications.

\section{Conclusions}

In this paper, we used polytropes to understand the fundamental properties of quasi-toroidal modes for slowly and uniformly rotating stars. 

The sectoral r modes of zero radial order are qualitatively different from the other r modes in that they do not rely on non-adiabatic stratification to balance the radial component of the Coriolis force. This is critical when the stratification is close to neutral (for polytropes, in the neighbourhood of $n=1.5$). In this neighbourhood, the modes with $\ell\neq m$ or $k\neq0$ all involve large radial displacements and are no longer nearly toroidal. The $\ell=m$, $k=0$ modes retain a small radial displacement through the star (except possibly at the surface if the surface has $p=\rho=T=0$ as pointed out by \citetalias{Provost}). The depth dependence of the horizontal displacement is close to $x^m$, as it is the only solution allowing both horizontal and vertical force balance, in the absence of viscosity and the lack of a buoyant contribution from non-adiabatic stratification.

\begin{figure}
    \centering
    \begingroup
        \sbox0{\includegraphics{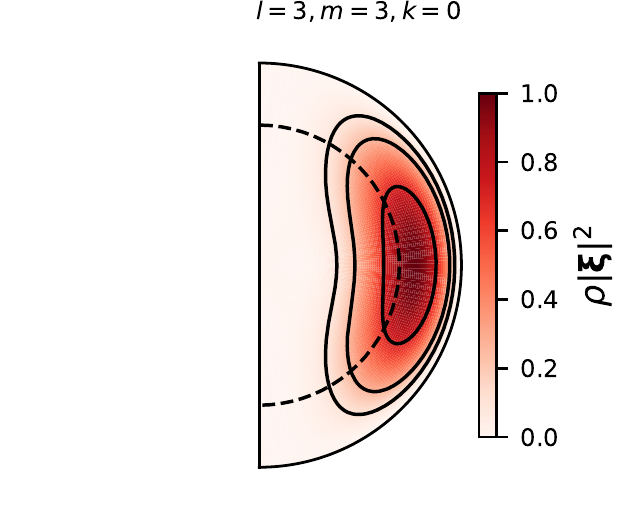}}%
        \includegraphics[clip,trim={.38\wd0} 0 0 0,width=0.45\columnwidth]{Fig7_n=149_l=3_m=3_k=0.pdf}
    \endgroup \quad
    \begingroup
        \sbox1{\includegraphics{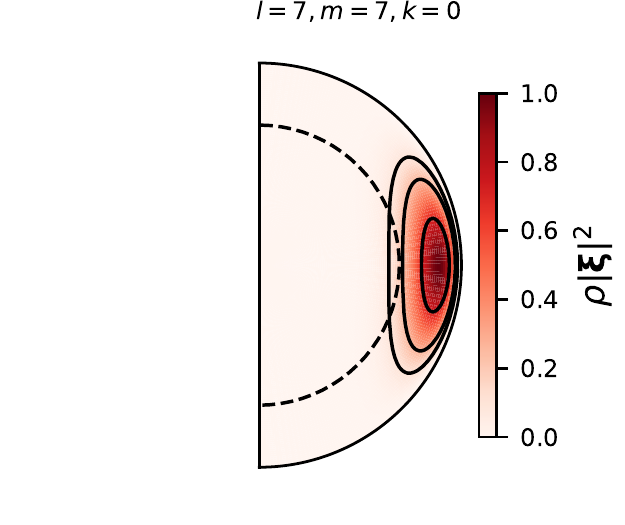}}%
        \includegraphics[clip,trim={.38\wd1} 0 0 0,width=0.45\columnwidth]{Fig7_n=149_l=7_m=7_k=0.pdf}
    \endgroup \\
        \begingroup
        \sbox2{\includegraphics{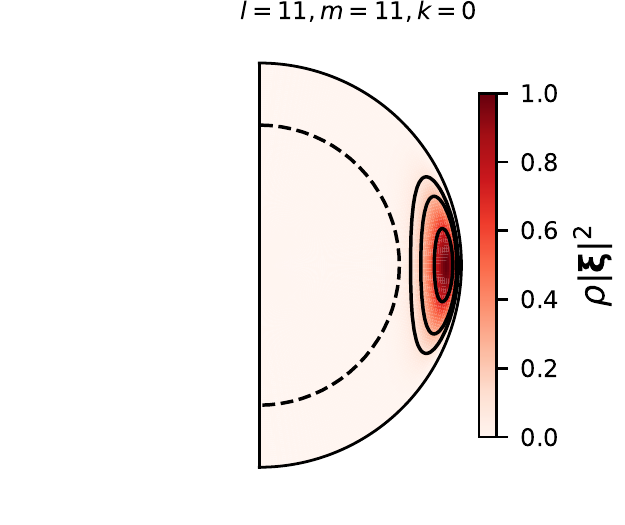}}%
        \includegraphics[clip,trim={.38\wd2} 0 0 0,width=0.45\columnwidth]{Fig7_n=149_l=11_m=11_k=0.pdf}
    \endgroup \quad
        \begingroup
        \sbox3{\includegraphics{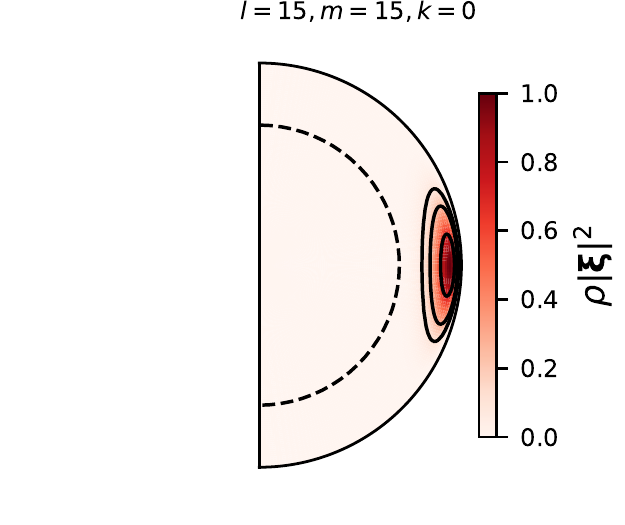}}%
        \includegraphics[clip,trim={.38\wd3} 0 0 0,width=0.45\columnwidth]{Fig7_n=149_l=15_m=15_k=0.pdf}
    \endgroup 
   \caption{Meridional cuts $(x,\theta)$ showing the normalised kinetic energy density of the sectoral modes of zero radial order, for $n=1.49$ and solar rotation, for different values of $m$ as labelled. The dashed line indicates the position of the base of the convection zone in the Sun, at $x=0.7$. The contour levels are marked by black lines in the colour bar.}
        \label{EKpol}%
    \end{figure}

\begin{figure}
    \centering
    \includegraphics[width=0.98\columnwidth]{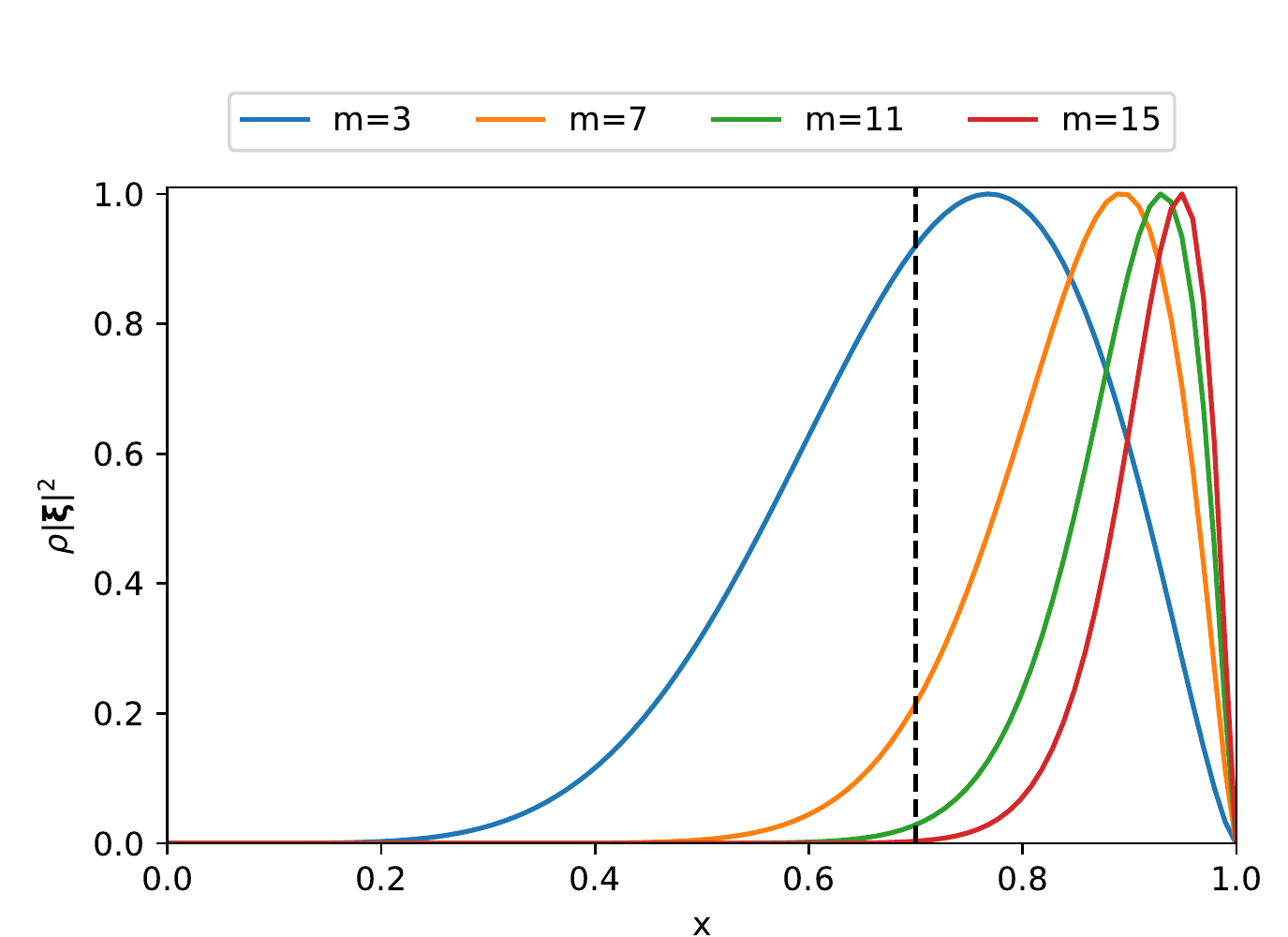}
   \caption{Radial dependence at the equator of the normalised kinetic energy density of the sectoral modes of zero radial order shown in Fig.~\ref{EKpol}. The dashed line indicates the position of the base of the convection zone in the Sun, at $x=0.7$.}
        \label{EKeq}%
    \end{figure}

Consequently, in the case of the Sun, we speculate that only the $\ell=m$, $k=0$ quasi-toroidal modes can exist in the convection zone, which is very close to adiabatically stratified (corresponding approximately to the polytropic index $n=1.49$). They are presumably the modes that are observed at the solar surface. Figure ~\ref{EKpol} shows the kinetic energy density associated with those modes on a meridional cut. In Fig.~\ref{EKeq}, we plot the corresponding radial dependence at the equator. These figures suggest that solar Rossby waves have diagnostic potential because different modes have different radial and latitudinal distribution of the kinetic energy density. It can be noticed that the kinetic energy density of the modes peaks in the interior (not at the surface as p modes would), and that the $m=3$ mode has a kinetic energy density that peaks near $x=0.75$. 

However the results of this study cannot be directly applied to the solar case, since the Sun is not a uniformly rotating inviscid polytrope. In particular, latitudinal differential rotation will modify the latitudinal eigenfunctions  \citep[][in prep.]{GizoninPrep}. Also, the inclusion of radial differential rotation has not been considered here. Furthermore, the Sun has a convectively stable radiative interior, beneath its convective envelope that requires a careful consideration of the matching conditions.

%-------------------------------------------------------------------
\begin{acknowledgements}
       We thank the referee for useful comments which helped improve the manuscript. We acknowledge partial support from the German Aerospace Center (Deutsches Zentrum für Luft und Raumfahrt) under PLATO Data Center grant 50OO1501. We also acknowledge partial support from ERC Synergy Grant WHOLESUN 810218. This research made use of NumPy \citep{van2011numpy}, SciPy \citep{jonesscipy2001} and matplotlib, a Python library for publication quality graphics \citep{Hunter:2007}. 
\end{acknowledgements}
%-------------------------------------------------------------------
 \bibliographystyle{aa} 
 \bibliography{bibrossby}
\end{document}